\documentclass[preprint,prl]{revtex4}

\usepackage{graphicx}
\usepackage{dcolumn}
\usepackage{bm}

\begin{document}

\setlength{\unitlength}{1mm}
\textwidth 15.0 true cm
\textheight 22.0 true cm
\headheight 0 cm
\headsep 0 cm
\topmargin 0.4 true in
\oddsidemargin 0.25 true in

\title{Post-Inflationary Curvature Perturbations}

\author{Andrei Gruzinov}
 \affiliation{Center for Cosmology and Particle Physics, Department of Physics, New York University, NY 10003}

\date{January 19, 2004}

\begin{abstract}

Curvature perturbations generated after the end of inflation are estimated for a generic multi-field inflationary potential. The curvature perturbations are generated  by conversion of inflationary isocurvature perturbations, similar to curvaton and inhomogeneous reheating scenarios. During inflation, the trajectory along which the fields are rolling down will fluctuate. When inflation ends, the fields enter the oscillatory phase along different directions in different parts of the universe. Different oscillatory trajectories can have different effective equations of state and reheating rates, so that different parts of the universe will evolve differently. This translates the isocurvature perturbations into the curvature perturbations. 

If all light fields make similar contributions to the potential energy, the postinflationary curvature perturbations are much smaller than the standard inflationary curvature perturbations. But if different light fields make very different contributions to the potential energy, the postinflationary curvature perturbations can exceed the inflationary perturbations. The characteristic feature of postinflationary perturbations is non-Gaussianity $\sim \pm 1$.

\end{abstract}

\pacs{}

\maketitle

\section{Introduction}

Standard one-field models of inflation \cite{standard1,standard2} have the obvious virtue of simplicity, and should be preferred to multi-field models, but only so long as no important effects are missed. It has been argued that such effects might indeed exist \cite{lyth,finelli,bern,DGZ1,DGZ2,Z,kofman}. 

An example relevant for this work is the ``inhomogeneous reheating'' scenario \cite{DGZ2}. According to the inhomogeneous reheating model, curvature perturbations are generated by conversion of inflationary entropy fluctuations in the following way. In addition to the inflaton field $\phi$ , there exists another light scalar field $\chi$ which has a very small potential energy. This field will fluctuate during inflation. After inflation, $\chi$ modulates the reheating rate, thus generating postinflationary curvature perturbations, which can be larger than the curvature perturbations generated during inflation.

This mechanism has obvious generalizations \cite{DGZ2}. Postinflationary curvature fluctuations can be generated by inhomogeneous reheating, or freeze-out, or mass domination, etc. The most important question is how these postinflationary curvature perturbations compare to the standard inflationary curvature perturbations.

We show that (i) Effects analogous to inhomogeneous reheating are present in generic multi-field inflation. (ii) ``Generically'' the postinflationary curvature perturbations are smaller than inflationary. (iii) Yet simple multi-field models with dominant postinflationary perturbations do exist. (iv) The characteristic feature of the postinflationary curvature perturbations is non-Gaussianity $\sim \pm 1$.

We describe the basic mechanism for generating curvature perturbation after inflation and estimate the size of the effect in \S 2. We describe simple models with dominant postinflationary curvature perturbations in \S 3, where we also estimate the non-Gaussianity of postinflationary curvature perturbations.

\section{The mechanism for generating postinflationary curvature perturbations and the estimate of the amplitude.}

After the end of inflation the inflaton fields $\phi _a$ will oscillate. Consider the potential energy in the region of oscillations. There can be directions in the $\phi _a$ space along which the potential is nearly quadratic. Oscillations along quadratic directions are equivalent to matter, with the equation of state $p=0$. There can be other directions in the $\phi _a$ space along which the potential is nearly quartic. Oscillations in the quartic potential are equivalent to radiation, with the equation of state $p=\rho /3$. If different parts of the universe enter the oscillation region along different directions, the difference in equations of state will be translated into curvature perturbations. In a similar way, different oscillatory trajectories can give rise to different reheating rates. 

A rough estimate of the resulting superhorizon curvature perturbation $\zeta_{\rm post}$ should be \begin{equation}\label{zeta-theta}
\zeta _{\rm post} \sim \delta \theta ,
\end{equation}
where $\theta$ are angles in the $\phi$ space. A rigorous definition of $\zeta$ is this: on superhorizon scales, $e^\zeta$ is proportional to the local scale factor on uniform energy density hypersurfaces. 

Now we need to estimate the superhorizon fluctuations in the directions $\delta \theta$ along which the fields enter the oscillation region. The field fluctuations during inflation are $\delta \phi \sim H$, where $H$ is the Hubble parameter during inflation. Then the fluctuations in the angle are 
\begin{equation}\label{theta}
\delta \theta \sim {H\over \phi}.
\end{equation}

From (\ref{zeta-theta}) and (\ref{theta}) we get an estimate for the postinflationary curvature perturbation, 
\begin{equation}\label{postpert}
\zeta _{\rm post} \sim {H\over \phi}.
\end{equation}
This should be compared to the standard inflationary curvature perturbation
\begin{equation}\label{infpert}
\zeta _{\rm infl} \sim {H\over \epsilon ^{1/2} M_{Pl}},
\end{equation}
where $2\epsilon \equiv (M_{Pl}V'/V)^2$ is the slow-roll parameter. The ratio of postinflationary to inflationary perturbations is then
\begin{equation}\label{ratio}
{\zeta _{\rm post} \over \zeta _{\rm infl}} \sim {\epsilon ^{1/2} M_{Pl}\over \phi}\sim {1\over N},
\end{equation}
where $N\gg 1$ is the number of e-foldings until the end of inflation. 

Thus the postinflationary curvature perturbations are ``generically'' subdominant. It is clear that one can cook up a potential energy which will give rise to a very strong conversion of entropy into curvature perturbations after the end of inflation. Or one can have a potential energy which gives rise to very large entropy perturbations, leading to postinflationary curvature perturbations greater than the standard ones. But, as the next section shows, this can actually occur in very simple two-field models. 

There also exists a radical way to ensure that the postinflationary curvature perturbations dominate. One can suppress the perturbations produced during the inflationary stage, by assuming that inflation occurs in a false vacuum state \cite{dvali}. This assumption also allows for a large level of gravitational waves in postinflationary perturbation models \cite{riotto}.

\section{Large postinflationary curvature perturbations}

If all fields give similar contributions to the potential energy, the postinflationary curvature perturbations are subdominant, as we have seen.  However, when this condition is violated, and different fields contribute very differently to the potential energy, the postinflationary curvature perturbations can dominate. 

We will give several examples of models with large post-inflationary curvature perturbations. For simplicity, we will assume that the parameters of the models are such that the postinflationary perturbations are the much larger than inflationary. We will therefore neglect the curvature perturbations generated during inflation. We want to demonstrate two things: (i) large post-inflationary curvature perturbations are not exotica, some simple models give large post-inflationary perturbations, (ii) the characteristic feature of the postinflationary perturbations is non-Gaussianity $\sim \pm 1$.   

The non-Gaussianity $\sim \pm 1$ is much larger than the standard one-field inflationary non-Gaussianity ($\sim$ tilt, \cite{maldacena}), and it might be observable in future galaxy surveys \cite{sco, note}.

\subsection{Inhomogeneous reheating I}
In inhomogeneous reheating \cite{DGZ2} the two fields play a very different role simply by assumption. It is postulated that there is an iflaton $\phi$ field and another light field $\chi$. The field $\chi$ has no significant energy density, yet it can modulate the rate of reheating. 

Assume that (i) reheating occurs during the time when $\phi$ oscillations behave like matter, (ii) reheating rate $\Gamma$ is proportional to $\chi ^2$, because $\chi$ is the coupling strength of the inflaton to radiation. 

Before reheating, the energy density $\rho \propto a^{-3}$, where $a$ is the scale factor.  The energy density at reheating is 
\begin{equation}
\rho \propto H^2\sim \Gamma ^2\propto \chi ^4,
\end{equation}
and the scale factor at reheating is
\begin{equation}
a_r\propto \chi ^{-4/3},
\end{equation}
After reheating the energy density is $\propto a_ra^{-4}\propto \chi ^{-4/3}a^{-4}$, and the perturbations of the scale factor on uniform energy hypersurfaces are 
\begin{equation}\label{a1}
e^{\zeta }\propto \chi ^{-1/3},
\end{equation}
or
\begin{equation}
\zeta = -{1\over 3} \ln \chi - const~=~-{1\over 3}\left( \delta _{\chi }-{1\over 2}\delta _{\chi }^2 \right),
\end{equation}
where $\delta _{\chi }\equiv \chi / <\chi> -1$ is the approximately Gaussian inflationary fluctuation. The definition of primordial non-Gaussianity $f_{nl}$ used by \cite{komsper, maldacena} is
\begin{equation}
\zeta = g-(3/5)f_{nl}g^2,
\end{equation}
where $g$ is Gaussian. We therefore get
\begin{equation}
f_{nl}=-{5\over 2}
\end{equation}
for this version of inhomogeneous reheating.

\subsection{Inhomogeneous reheating II}
The version of inhomogeneous reheating presented above requires a non-renormalizable coupling, $\propto \phi \chi \bar{\psi }\psi $, where $\phi$ is the inflaton, $\chi$ is the light scalar field modulating the rate of reheating, and $\psi$ is the fermion field. The resulting reheating rate is then $\Gamma \propto \chi ^2$.

One can get the same rate of reheating in a different way. Let there be two inflaton fields $\phi _1$ and $\phi _2$ with different masses. Assume that the scalar field $\chi$ modulates the rate of $\phi _1-\phi _2$ conversion via a renormalizable coupling $\propto \chi \phi _1 \phi _2$. Further assume that $\phi _2$ reheats much faster than $\phi _1$. For small $\chi$, the effective reheating rate will be determined by the admixture of the $\phi _2$ ``flavor'' in the $\phi _1$ mode. The latter scales as $\phi _2\propto \chi \phi_1$. The resulting reheating rate is again $\Gamma \propto \chi ^2$, and the results of the previous section apply.

\subsection{``Curvaton''}
Consider a two-field model with potential energy
\begin{equation}
V(\phi , \chi ) = \lambda \phi ^4 + m^2\chi ^2.
\end{equation}
Since now the two fields give very different contributions to the potential energy, the postinflationary perturbations can be significant. Indeed, for some values of parameters, the model reduces to the curvaton \cite{lyth}. We get a curvaton if the initial conditions are such that (i) $\phi$ is the inflaton field, it carries the energy during  inflation, (ii) $\chi$ is light during inflation, so that it fluctuates. 

After the end of inflation, the energy stored in $\phi$ oscillations will redshift away, and only the energy in $\chi$ oscillations will remain. This will translate the $\chi$ perturbations into the curvature perturbations. The energy density of the $\chi$ field will be $\propto \chi ^2a^{-3}$, where $\chi$ is the value at horizon crossing. This gives the curvature perturbation 
\begin{equation}\label{a2}
e^{3\zeta }\propto \chi ^2.
\end{equation}
The corresponding non-Gaussianity is 
\begin{equation}
f_{nl}=5/4.
\end{equation}

\subsection{Inhomogeneous reheating III}
Consider inhomogeneous reheating of the iflaton $\phi$. The field $\chi$ which modulates the rate of reheating is now assumed to have a small mass. After reheating the field $\chi$ will redshift slower than radiation, and therefore it will generate the curvaton-like perturbations, which will add (coherently) to the perturbations due to inhomogeneous reheating. As a result, we can calculate the perturbations by multiplying equations (\ref{a1}, \ref{a2})
\begin{equation}
e^{3\zeta }\propto \chi ^2~\chi ^{-1}~=~\chi
\end{equation}
and the corresponding non-Gaussianity is 
\begin{equation}
f_{nl}=5/2.
\end{equation}

The detailed calculation can be done as follows. After the end of inflation, the inflaton oscillations redshift like matter. The Hubble parameter $H$ will be given by 
\begin{equation}
H^2\propto a^{-3},
\end{equation}
where $a$ is the scale factor. Reheating occurs when $H\sim \Gamma \propto \chi ^2$. Thus the scale factor at reheating
\begin{equation}
a_r\propto \chi ^{-4/3},
\end{equation}
here and in what follows $\chi$ denotes the value of the $\chi$ field at the moment of horizon crossing. The Hubble parameter at reheating 
\begin{equation}
H^2_r\propto a_r^{-3}.
\end{equation}
After reheating,
\begin{equation}
H^2\propto H^2_ra_r^4a^{-4}\propto a_ra^{-4}
\end{equation}
The $\chi$ field will start to oscillate when $H^2\sim m_{\chi }^2$. This occurs at the scale factor
\begin{equation}
a_{\chi }\propto a_r^{1/4}.
\end{equation}
After that, the energy density of the $\chi$ field will scale as 
\begin{equation}
\rho  _{\chi }\propto \chi ^2a_{\chi }^3a^{-3}.
\end{equation}
Since the $\chi$ energy will ultimately dominate, the fluctuations of the scale factor on uniform energy hypersurfaces are given by
\begin{equation}
e^{3\zeta }\propto \chi ^2a_{\chi }^3\propto \chi ^2a_r^{3/4}\propto \chi ,
\end{equation}
as expected.

\subsection{Non-Gaussianity}
One can keep on going and invent other models with large postinflationary curvature perturbations. It appears that in all such models, after the conversion of entropy into curvature perturbations, and  after the final reheating of the massive fields, the radiation energy density becomes proportional to a certain power of some light scalar field $\chi$. Thus, quantum fluctuations of $\chi$ which occur during the inflationary stage are translated, after the end of inflation, into curvature perturbations
\begin{equation}
\zeta \sim \ln (1+ \delta _{\chi }).
\end{equation}
If these postinflationary perturbations dominate, the non-Gaussianity is 
\begin{equation}
f_{nl}\sim \pm 1.
\end{equation}
The previous sections give examples with $f_{nl}=-2.5,~~1.25,~~2.5$.

\section{Conclusion}

\begin{itemize}

\item Effects analogous to inhomogeneous reheating are present in generic multi-field inflation. 
\item Generically the postinflationary curvature perturbations are smaller than inflationary. 
\item Yet simple multi-field models with dominant postinflationary perturbations do exist. 
\item The characteristic feature of the postinflationary curvature perturbations is the relatively large non-Gaussianity $\sim \pm 1$.

\end{itemize}

\begin{acknowledgments}
I thank Gia Dvali and Gregory Gabadadze for useful discussions. This work was supported by the David and Lucile Packard Foundation.
\end{acknowledgments}


\begin{thebibliography}{}
\bibitem{note}
One, of course, should first translate the primordial (inflationary or postinflationary) non-Gaussianities to the non-Gaussianities of the modes after re-entering the horizon \cite{gruzinov}.
\end{thebibliography}
\end{document}